\documentclass[aps,pra,floatfix,amsmath,amssymb,showpacs,showkeys,twocolumn,10pt]{revtex4-1}
\usepackage[caption=false]{subfig}
\usepackage{graphicx,bm,color}
\usepackage{amsfonts}
\usepackage{epstopdf}
\begin{document}

\frenchspacing

\title{Engineering of orbital angular momentum spectrum of down-converted photons with mode-invariant pump}

\begin{abstract}
In this article we investigate the orbital angular momentum spectrum of photons produced in parametric down conversion process. We study how the transverse profile of the pump mode affects the spectrum as compared to a gaussian pump beam. We show that using perfect optical vortex mode as pump the spectrum can be highly narrowed down. We further study the entanglement between the two down converted photons in the orbital angular momentum basis. We observe that photons entangled in Laguerre-Gaussian modes exhibit greater higher dimensional entanglement.
\end{abstract}

\author{Anindya Banerji}\email[Electronic address: ]{abanerji09@gmail.com}
\author{Ali Anwar}
\author{Hrushikesh Sable}
\author{Nijil Lal}
\author{Ravindra P. Singh}
\affiliation{Physical Research Laboratory, Ahmedabad 380009, India}

\date{\today}

\maketitle

\section{Introduction}

The method of spontaneous parametric down conversion or SPDC in short has been a long standing method for the generation of entangled photons \cite{Kwiat}. In this method, a pump photon at a higher frequency is absorbed in a nonlinear crystal and two photons at lower frequencies are emitted. The emitted photons obey the energy and momentum conservation law which leads to the following
\begin{subequations}
\begin{align}
\label{PhaseMatching1}
\hbar\omega_p=\hbar\omega_s+\hbar\omega_i \\
\label{PhaseMatching2}
\hbar\vec{k}_p=\hbar\vec{k}_s+\hbar\vec{k}_i
\end{align}
\end{subequations}

\noindent where the subscripts $p$, $s$ and $i$ denote $pump$, $signal$ and $idler$, respectively. Under degenerate phase matching conditions, the emiited photons have identical angular frequencies $\left(\omega_s=\omega_i=\omega_p/2\right)$. In this situation, photons sampled from diametrically opposite points on the emission cone (assuming noncollinear phase matching) exhibit maximum correlations in various degrees of freedom (DOF). These correlations arise from conservation principles of respective DOFs and have been verified to give rise to entanglement between the two photons \cite{Hong, Walborn, Burnham, Ghosh, Ou}. The presence of these correlations have led to the study of higher dimensional entanglement \cite{Dada} as well as the generation and study of hyperentangled \cite{Barreiro,Barbieri} and hybrid-entangled states \cite{Neves} . In higher dimensional entanglement, the pair of photons are entangled in such a DOF that has an infinite dimensional Hilbert space associated with it. One such DOF is the orbital angular momentum (OAM) \cite{Andrews}.\\
The OAM degree of freedom has been subjected to intense scrutiny over the years \cite{Krenn}. Light beam with OAM is most commonly defined by the Laguerre-Gauss (LG) modes \cite{Allen} which are exact solutions of the paraxial wave equation in cylindrical coordinates and forms a complete basis. Such modes are characterised by a well-defined phase structure and spatially varying amplitude. The conservation of OAM in the process of SPDC was experimentally demonstrated by Mair et al \cite{Mair}. They also confirmed that besides conservation, the signal and idler photons are also entangled in OAM. This gave rise to an entire array of literature that studied the use of OAM as a basis in various quantum information protocols. One of the reasons for such huge interest in the field was the dimension of the associated Hilbert space. In principle $l$ can range from $-\infty$ to $+\infty$, effectively giving rise to an infinite dimensional Hilbert space. It was realised that $d$-dimensional quantum states or \emph{qudits} can be implemented using the OAM basis \cite{Vaziri,Heismayr}. Amidst all this development, the study of the OAM spectrum of the signal and idler photons assumed a lot of importance specifically from the point of generation of higher dimensional entangled states. But majority of the articles have only looked at the spectrum for either a gaussian pump mode \cite{Miatto} or LG pump modes \cite{Arnaut,Yao, Osorio}. Recently, it has been shown that any arbitrary spatial mode of the pump can be directly transferred to the signal and idler photons \cite{Uren,Ali}. Therefore, it becomes interesting to study how different spatial modes affect the OAM spectrum. We aim to address this issue with our article.\\
Here, we compare the OAM spectrum of the SPDC photons for different spatial modes of the pump. We consider LG modes of varying order as the pump field and look at the spectrum when both the output photons are also projected onto LG modes. Further, we replace the LG mode in the pump with perfect optical vortex (POV) mode \cite{Vaity} and look at the OAM distribution in the signal and idler modes. A perfect optical vortex is characterised by an annular ring shaped transverse profile, the radius and width of which remains fixed for all OAM values. We also look at how the projection of the signal and idler modes to POV modes affects the spectrum when a POV pump is used. Further, we wish to address how the OAM spectrum for different modes affects the higher dimensional entanglement.\\
The paper is organised as follows. In section II, we have presented the mathematical background leading to the main results of this article. In section III, we have studied the azimuthal spectrum in detail for different pump beams. Section IV deals with the entanglement between the signal and idler photons as a function of the pump beam. We conclude this article with section V, where we review the main results of this article and discuss future applications and implications.

\section{Mathematical background}

\begin{figure}
\includegraphics[scale=0.4]{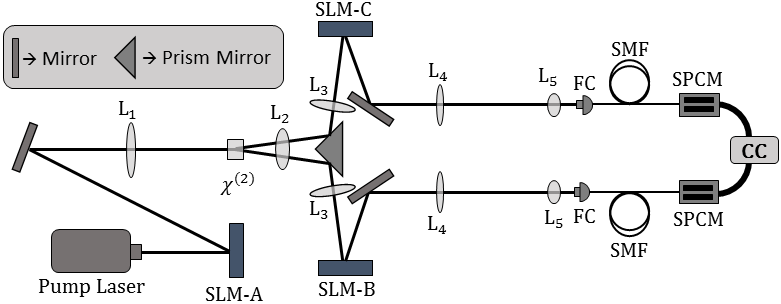}
\caption{Experimental schematic. The pump is prepared in any desired spatial mode by using the spatial light modulator (SLM-A). L1 is Fourier transforming lens that is used along with SLM-A to prepare the pump in POV modes. $\chi ^2$ is a nonlinear crystal where the downconversion takes place. A prism mirror is used deflect the signal and idler along different paths. SLM-B and SLM-C is used to project the signal and idler photons to differnt spatial modes which are then detected at the SPCM (single photon counting module). FC is fibre coupler, SMF is single mode fibre, CC is coincidence counter and L2-L5 are imaging lens.}
\label{fig:ExpScheme}
\qquad
\end{figure}

The SPDC process is generally represented in the interaction picture using the Hamiltonian
\begin{equation}
\label{SPDCHamiltonian}
H_I=\chi\int_Vd^3rE_{p}^{+}\left(r,t\right)\hat{E}_{s}^{-}\left(r,t\right)\hat{E}_{i}^{-}\left(r,t\right)
\end{equation}
\noindent where $\chi$ is the susceptibility of the crystal, $E_j$ is the electric field associated with the $j$-th mode. Signal and idler are the two output modes of the SPDC process. In a semi-classical approach, the pump field is considered strong compared to the signal and idler fields and treated classically. We write it as
\begin{equation}
\label{pump}
E_{p}^{+}\left(r,t\right)=\psi_p\left(r\right)\exp\left(i\left(k_pz-\omega_pt\right)\right)
\end{equation}
\noindent where $\psi(r)$ is the mode function governing the transverse field distribution of the pump mode. The signal and idler fields are treated as quantum mechanical operators. They can be written as
\begin{equation}
\label{signalidler}
\hat{E}_{j}^{-}\left(r,t\right)=a^{\dagger}_j\psi_j\left(r\right)\exp\left(i\omega_jt\right)
\end{equation}
\noindent where $j\in\lbrace s,i\rbrace$. $a^{\dagger}_j$ is the bosonic creation operator for the $j$-th mode. It creates a photon in the $j$-th mode with transverse mode function $\psi_j$ and energy $\hbar\omega_j$. The two photon wavefunction is then written as
\begin{equation}
\label{wavefunction}
\vert\psi\left(t\right)\rangle=-i\exp\left(\int dt'H\left(t'\right)\right)\vert 00\rangle
\end{equation}
\noindent which under first order expansion yields the form
\begin{equation}
\label{firstorder}
\vert\psi\left(t\right)\rangle=\int dt'H\left(t'\right)\vert 00\rangle
\end{equation}
The time integral can be solved by invoking the phase matching conditions of Eq. (\ref{PhaseMatching1}). It results in a factor that modulates the wavefunction in time. Since it does not affect the spatial distribution of the signal and idler modes, we scale this factor to unity without any loss of generality. Eq. (\ref{firstorder}) can then be written as
\begin{eqnarray}
\label{coeff1}
\vert \psi\left(t\right)\rangle &=& A\int_V d^3r E_{p}^{+}\left(r,t\right)\hat{E}_{s}^{-}\left(r,t\right)\hat{E}_{i}^{-}\left(r,t\right)\vert 00 \rangle \nonumber \\
&=& A\int_V d^3r \psi_p\left(r\right)\psi_s\left(r\right)\psi_i\left(r\right)a^{\dagger}_s a^{\dagger}_i\vert 00 \rangle \nonumber \\
&=& A\int_V d^3r \psi_p\left(r\right)\psi_s\left(r\right)\psi_i\left(r\right)\vert \psi_s,\psi_i \rangle
\end{eqnarray}
\noindent where $\vert \psi_s,\psi_i\rangle$ is the two photon quantum state of the signal and idler photon associated with the mode function $\psi_s\left(r\right)$  and $\psi_i\left(r\right)$ respectively and \emph{A} is some constant in arbitrary units. The spectrum for any given pump can thus be calculated from Eq. (\ref{coeff1}) by suitably modifying the mode functions to suit the purpose and calculating the overlap integral.

\section{OAM spectrum}
\begin{figure}
\subfloat[$l_p=0$]{
\includegraphics[scale=0.4]{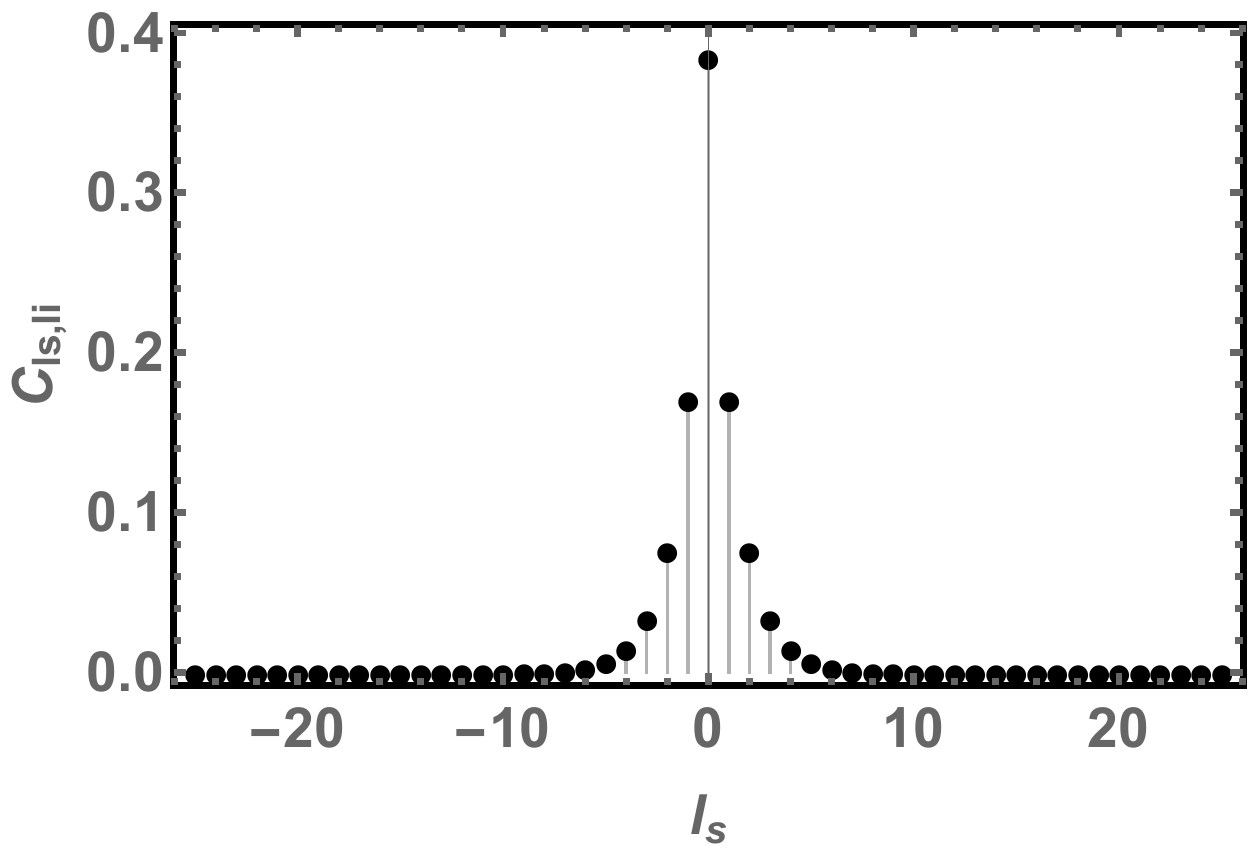}
\label{fig:label-e}}
\qquad
\subfloat[$l_p=1$]{
\includegraphics[scale=0.4]{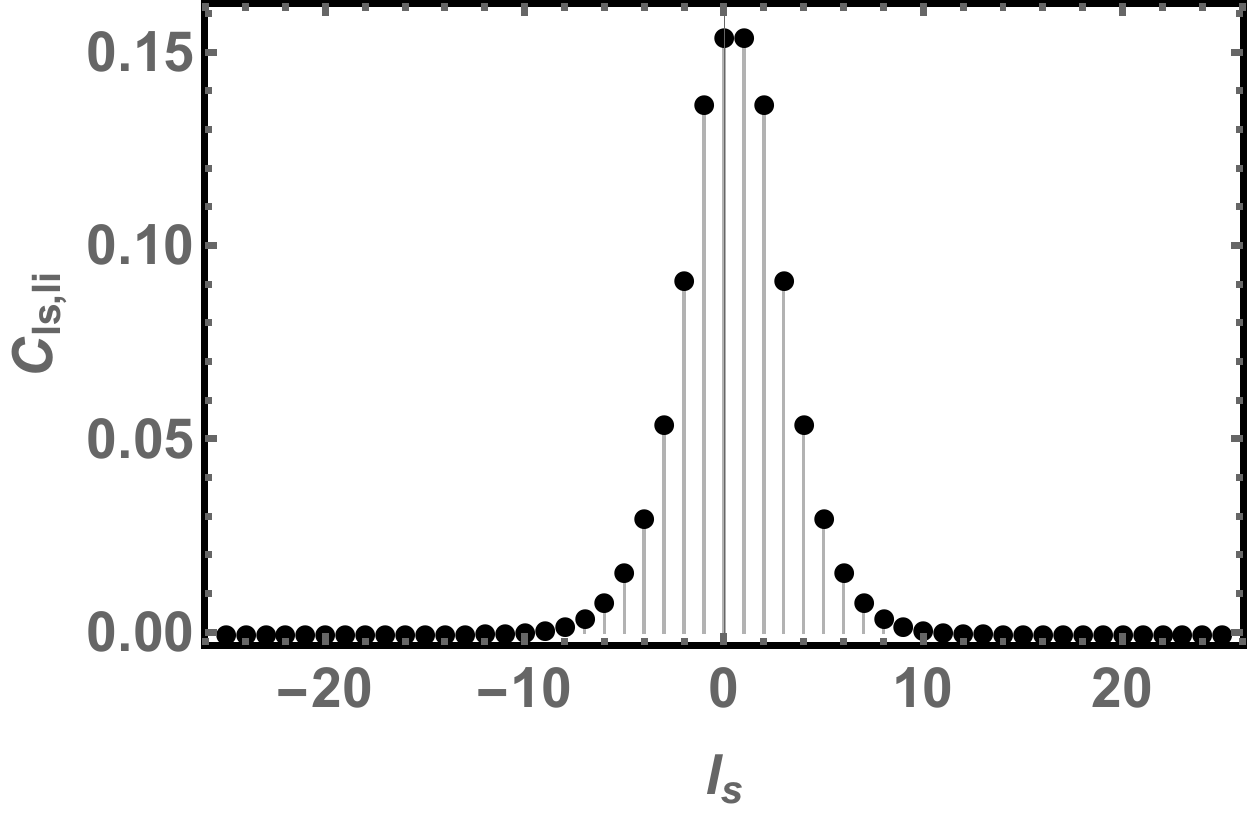}
\label{fig:label-f}}
\qquad\\
\subfloat[$l_p=2$]{
\includegraphics[scale=0.4]{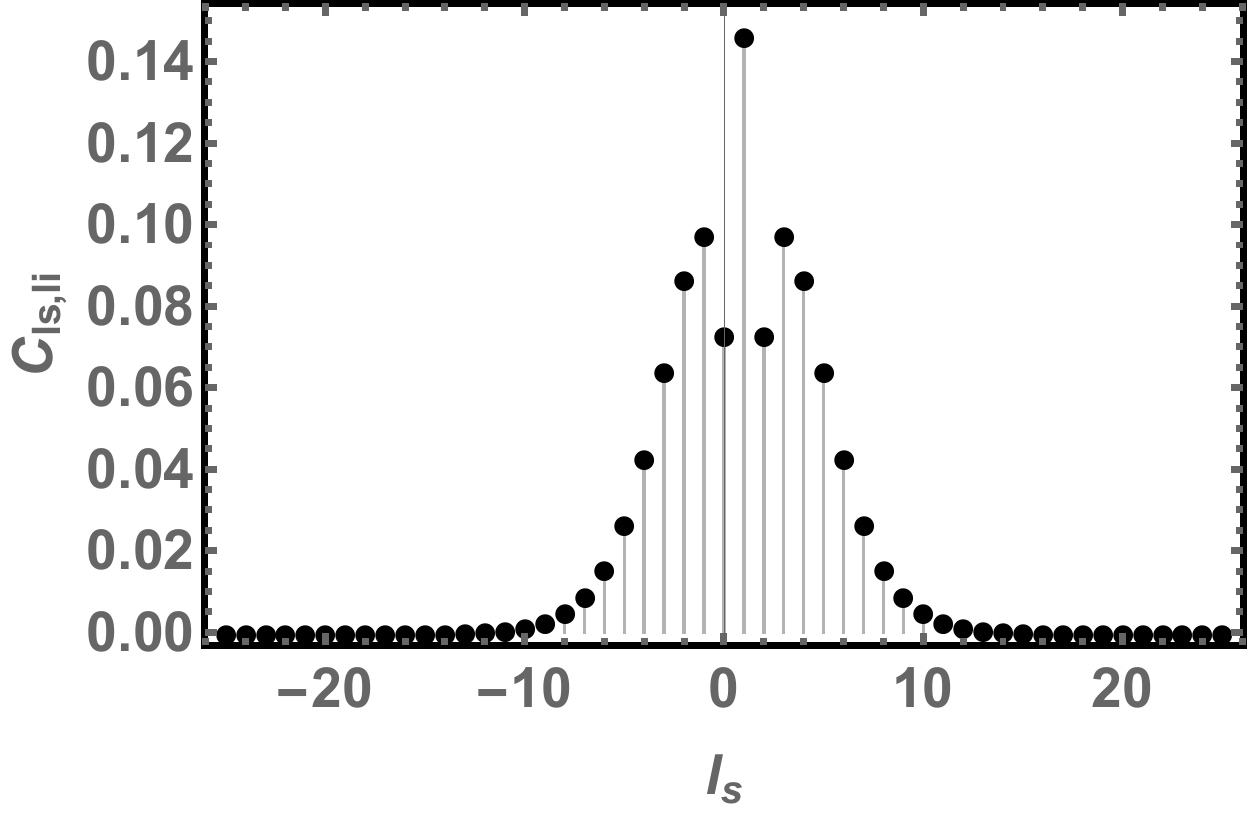}
\label{fig:label-g}}
\qquad
\subfloat[$l_p=3$]{
\includegraphics[scale=0.4]{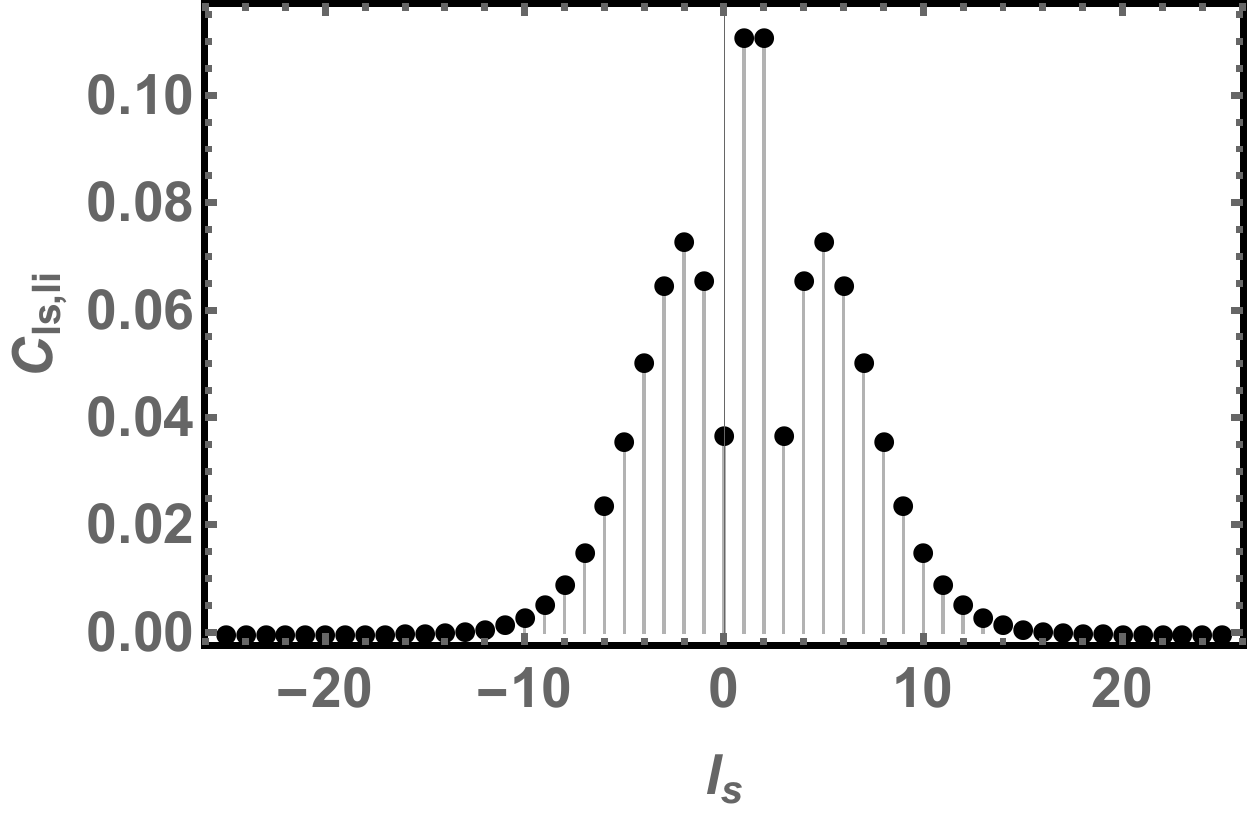}
\label{fig:label-h}}
\caption{OAM spectrum of the signal and idler photons for different pump OAM. All the three modes are LG modes.}
\label{fig:LGSpectrum}
\end{figure}
Single photons carrying orbital angular momentum are most commonly represented using Laguerre-Gauss mode functions. These mode functions are exact solutions of the paraxial wave equation in cylindrical coordinates $\left(\rho,\phi,z\right)$. In the most general case, they can be represented as
\begin{eqnarray}
\label{LGbeam}
\psi_{lp}\left(r,\phi,z\right) &=& \sqrt{\frac{2p!}{\pi\left(\vert l \vert +p\right)!}}\sqrt{\frac{1}{w(z)}}\left(\frac{\sqrt{2}r}{w(z)}\right)^{\vert l \vert}e^{-il\phi}\nonumber\\
&\times & L_p^{\vert l \vert}\left(\frac{2r^2}{w^2(z)}\right)\exp\left(-\frac{r^2}{w^2(z)}\right)
\end{eqnarray}
\noindent where $l$ is the azimuthal index number describing the helical structure of the photon, $p$ is the number of radial nodes, $w(z)=w_0\sqrt{1+z^2/z^2_R}$ is the gaussian beam width at distance $z$, $z_R$ is the Rayleigh length and $L_p^{\vert l \vert}(x)$ is the associated Laguerre polynomial which is given by
\begin{equation}
\label{LaguerrePolynomial}
L_p^{\vert l \vert}(x) = \sum_{m=0}^p \frac{\left(\vert l \vert +p\right)!}{(p-m)!(\vert l \vert +m)!m!}x^m
\end{equation}
Optical fields described by Eq. (\ref{LGbeam}) are also know as optical vortex due to the helical phase structure. The azimuthal index is then also referred to as the charge of the optical vortex. Now, looking at the $z=0$ plane and scaling the beam waist $w_0$ to unity in arbitrary units, Eq. (\ref{LGbeam}) can be simplified to
\begin{eqnarray}
\label{LGsimple}
\psi_{lp}\left(r,\phi\right) &=& \sqrt{\frac{2p!}{\pi\left(\vert l \vert +p\right)!}}\left(\sqrt{2}r\right)^{\vert l \vert} \nonumber\\
&\times & L_p^{\vert l \vert}\left(2r^2\right)\exp\left(-r^2\right)\exp\left(-il\phi\right)
\end{eqnarray}
The above equation describes a photon with a Laguerre-Gaussian transverse mode function. Now, the two photon output state of SPDC can be written in the OAM basis as follows
\begin{equation}
\label{OAMBasis}
\vert\psi\rangle=\sum_{l_s,l_i,p_s,p_i}C_{l_s,l_i}^{p_s,p_i}\vert l_s,p_s\rangle\vert l_i,p_i\rangle
\end{equation}
\noindent where $\vert l_j,p_j\rangle$ denotes a single photon in the $j$-th mode characterised by a LG transverse distribution with azimuthal index $l_j$ and radial index $p_j$. The azimuthal index runs from $-\infty$ to $\infty$ while the radial index runs from $0$ to $\infty$. The OAM space is thus an infinite Hilbert space. Qubits in the OAM basis can therefore be considered as qudits.\\
\begin{figure}[h!]
\subfloat[$l_p=0$]{
\includegraphics[scale=0.4]{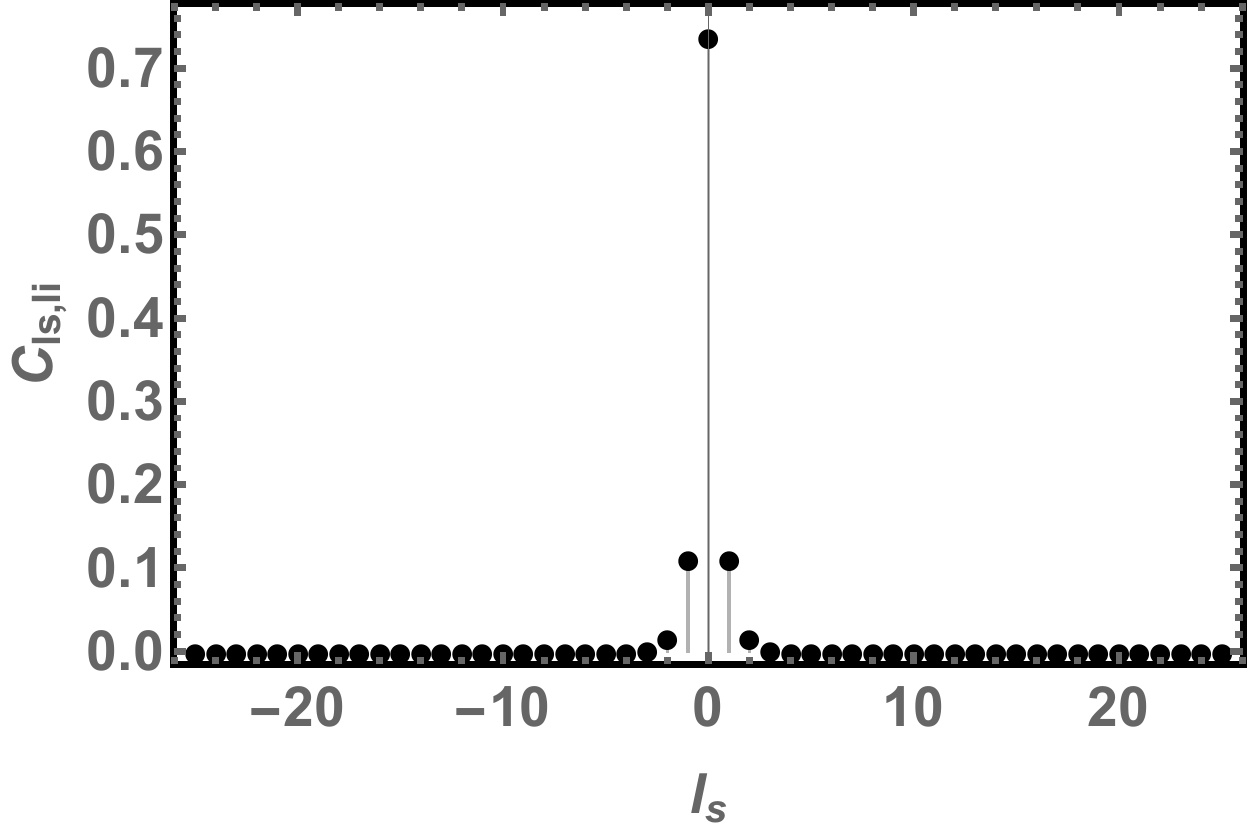}
\label{fig:label-e}}
\qquad
\subfloat[$l_p=1$]{
\includegraphics[scale=0.4]{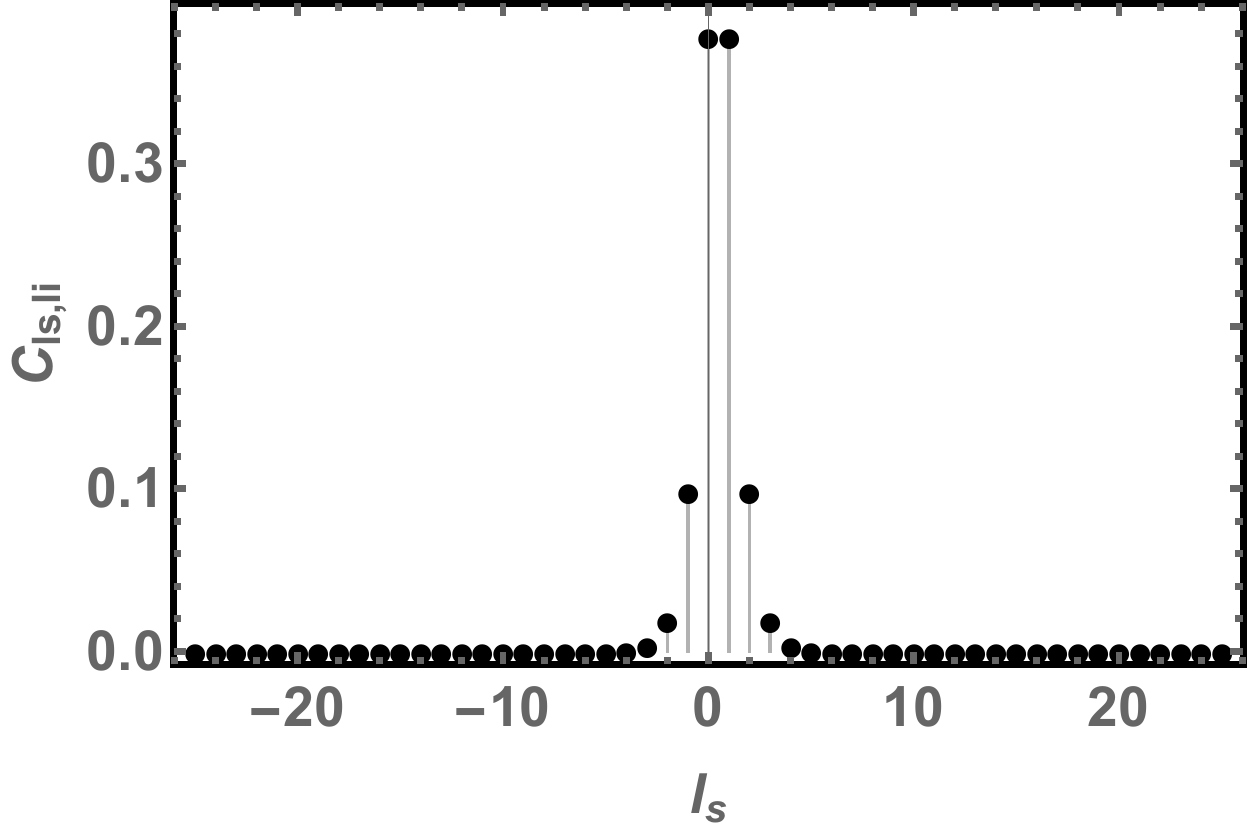}
\label{fig:label-f}}
\qquad\\
\subfloat[$l_p=2$]{
\includegraphics[scale=0.4]{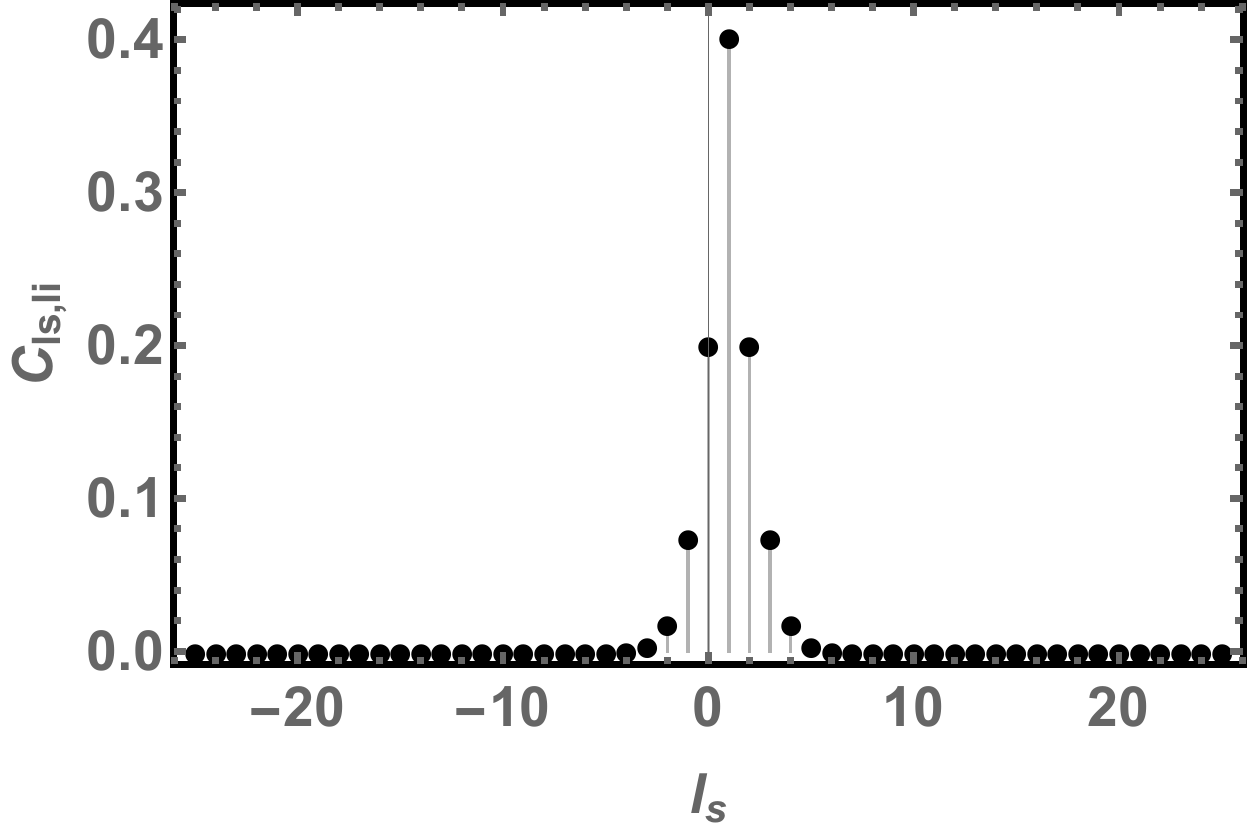}
\label{fig:label-g}}
\qquad
\subfloat[$l_p=3$]{
\includegraphics[scale=0.4]{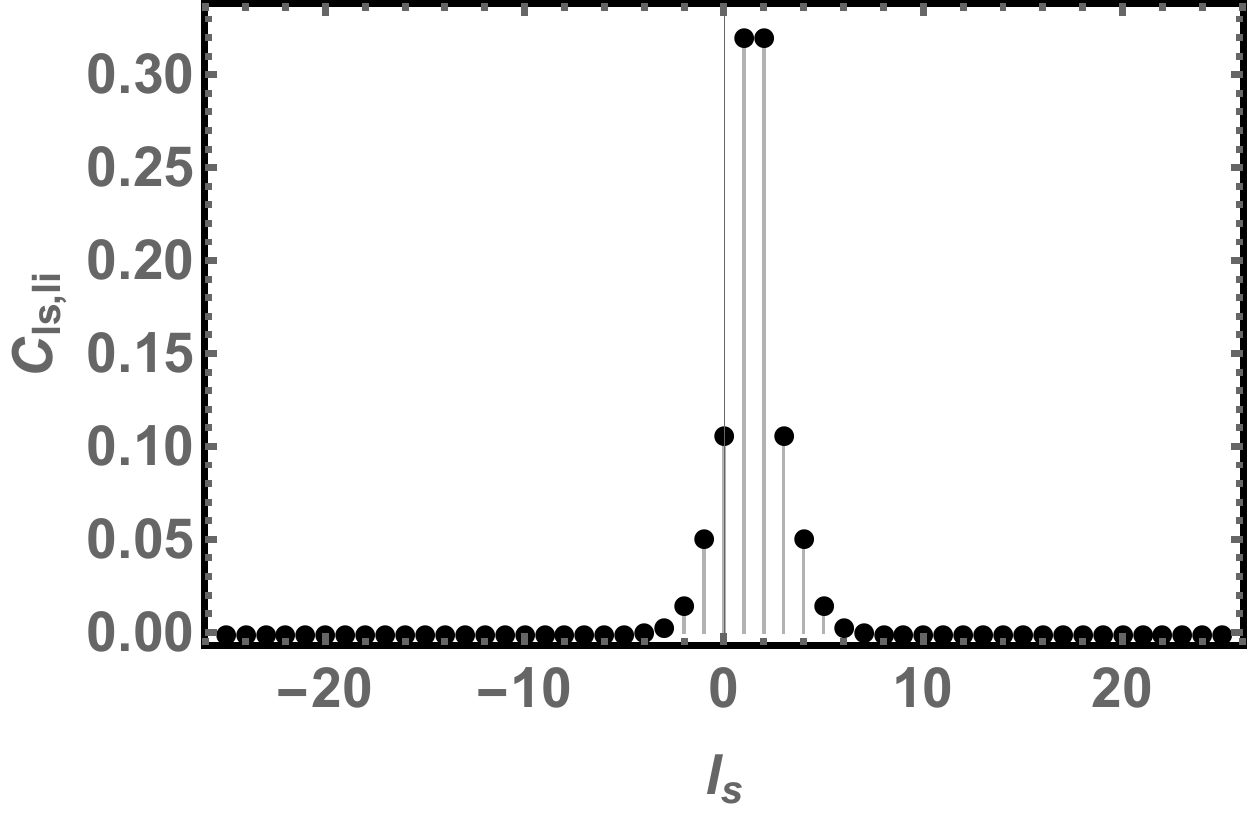}
\label{fig:label-h}}
\caption{OAM spectrum of the signal and idler photons for different pump OAM. Here the pump is in a POV mode and the signal and idler are projected to LG modes.}
\label{fig:PLGSpectrum}
\end{figure}
We are interested to find out how the azimuthal order $l_p$ of the incident pump is distributed between the signal and idler photons. In order to do so, we need to calculate the expansion coefficients $C_{l_s,l_i}^{p_s,p_i}$. Comparing with Eq. (\ref{coeff1}), it is easily seen that 
\begin{equation}
\label{coeff2}
C_{l_s,l_i}^{p_s,p_i} = \int_V d^3r~ \psi_p\left(r\right)\psi_s\left(r\right)\psi_i\left(r\right)
\end{equation}
\noindent where the mode functions are to be suitably replaced by LG mode functions as in Eq. (\ref{LGsimple}) with proper indices. Taking the radial index of the pump, signal and idler to be identically 0, we can write Eq. (\ref{coeff2}) as
\begin{eqnarray}
\label{coeffexpansion}
C_{l_s,l_i}&=&\sqrt{\frac{2^3}{\pi ^3\vert l_p\vert !\vert l_s\vert !\vert l_i\vert !}}2^{\frac{\vert l_p\vert +\vert l_s\vert +\vert l_i\vert }{2}}\nonumber\\
&\times &\int r dr d\phi ~r^{\vert l_p\vert +\vert l_s\vert +\vert l_i\vert}e^{-3r^2}e^{i\Delta l\phi}
\end{eqnarray}
\noindent where $\Delta l=l_p-l_s-l_i$. Since OAM is conserved in a down conversion process, $l_s$ and $l_i$ should always add up to $l_p$ in both magnitude as well as helicity. Using this criterion, the $\phi$ integral is evaluated to $2\pi$. The rest of the integral is evaluated using standard integration technique to the following
\begin{equation}
\label{finalcoeff}
C_{l_s,l_i} = P\left(\frac{2}{3}\right)^L\sqrt{\frac{1}{\vert l_p\vert !\vert l_s\vert !\vert l_i\vert !}}L!
\end{equation}
\noindent where $L=\left(\vert l_p\vert +\vert l_s\vert +\vert l_i\vert\right)/2$. $P$ is a constant independent of the azimuthal indices of any of the modes and is equal to $(1/3)\sqrt{8/\pi}$.\\
In Fig. (\ref{fig:LGSpectrum}), we study the OAM spectrum of the LG modes. For a pure gaussian mode ($l_p=0,r_p=0$), it is observed that the signal and idler modes always carry equal and opposite azimuthal charge. But the probability of both the photons being in a gaussian mode is maximum. So, the combined two photon state does not carry any net azimuthal charge. This arises solely from OAM conservation. In this case the output state of the down conversion process can be simply written as
\begin{equation}
\label{gaussian}
\vert \psi\rangle = \sum_{l=-\infty}^{\infty}C_l\vert l\rangle_s\vert -l\rangle_i
\end{equation}
\noindent This means, individually, each mode is in a thermal state of the OAM basis. In case of the pump field being an optical vortex of charge $1$, it is seen that the maximum probability shifts. The superposition state $\vert 1,0\rangle +\vert 0,1\rangle$ is the most probable state. An interesting pattern in the OAM spectrum is observed for all higher orders of the pump mode. For even orders, the superposition state $\vert l,0\rangle+\vert 0,l\rangle$ has the maximum probability where $l$ is the azimuthal index of the pump field. For odd orders, it is $\vert \frac{l+1}{2},\frac{l-1}{2}\rangle+\vert \frac{l-1}{2},\frac{l+1}{2}\rangle$ that has the maximum probability. Also, there is a secondary maxima for all higher orders of the pump field. This anomaly can be explained by further investigating Eq. (\ref{finalcoeff}). The factorial in the denominator is a decreasing function where as the factorial in the numerator is an increasing function. The secondary maxima occurs at those points where both the functions have similar weightage. This anomaly can be removed by using perfect optical vortex (POV) modes in place of LG modes.\\

\begin{figure}[h!]
\subfloat[$l_p=0$]{
\includegraphics[scale=0.4]{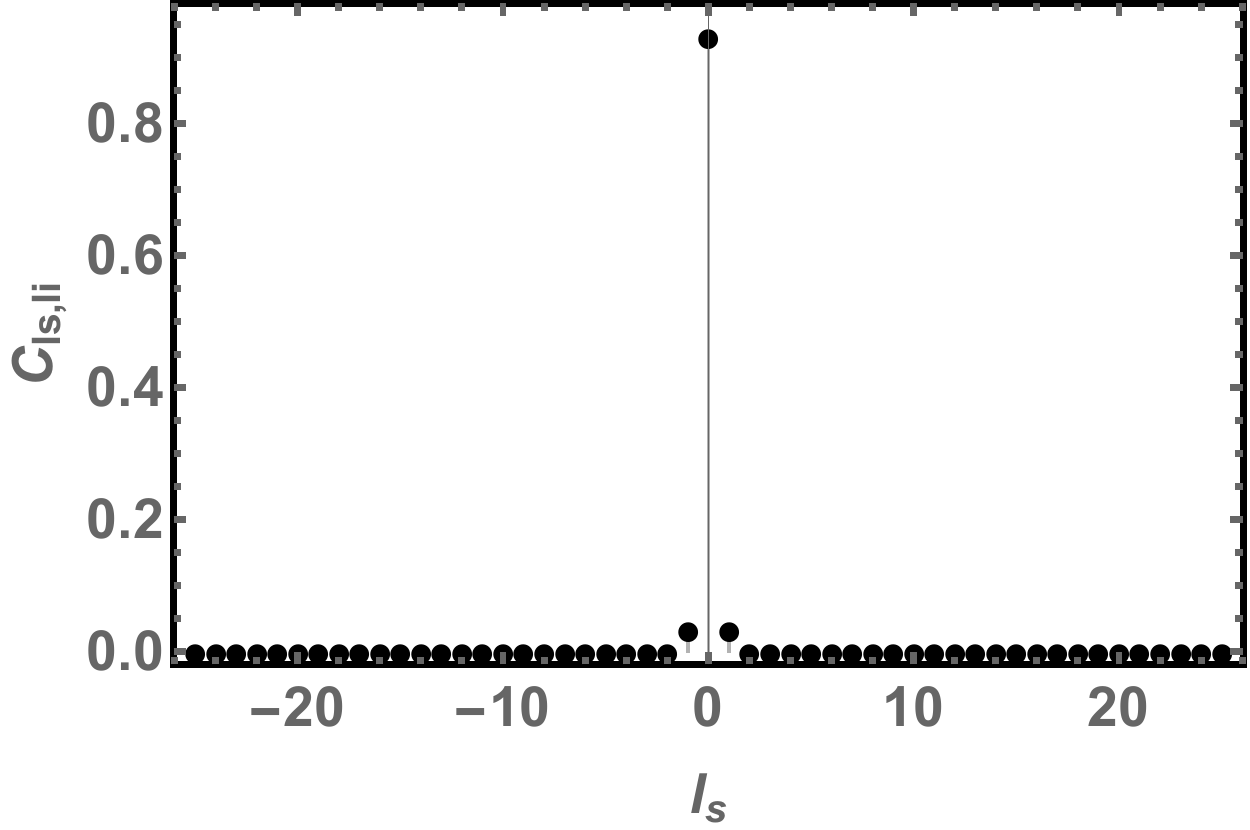}
\label{fig:label-e}}
\qquad
\subfloat[$l_p=1$]{
\includegraphics[scale=0.4]{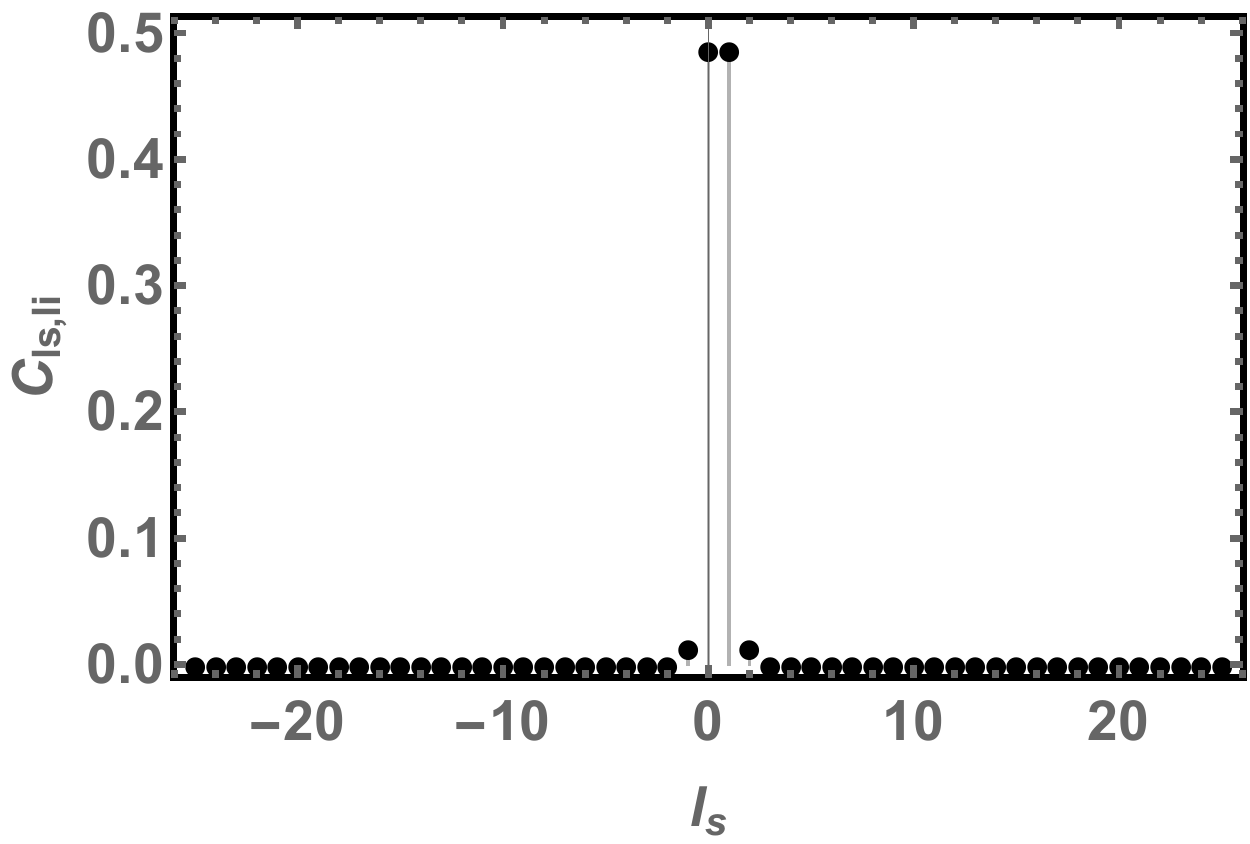}
\label{fig:label-f}}
\qquad\\
\subfloat[$l_p=2$]{
\includegraphics[scale=0.4]{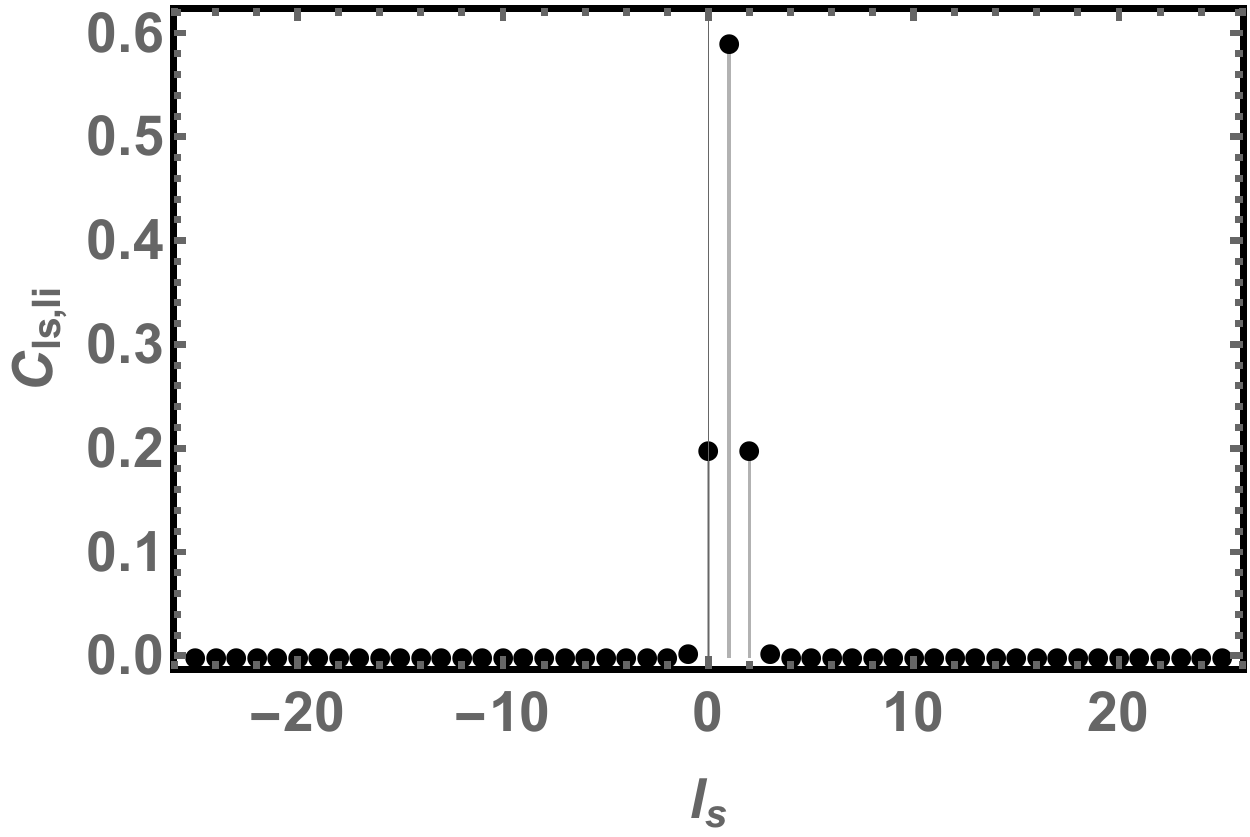}
\label{fig:label-g}}
\qquad
\subfloat[$l_p=3$]{
\includegraphics[scale=0.4]{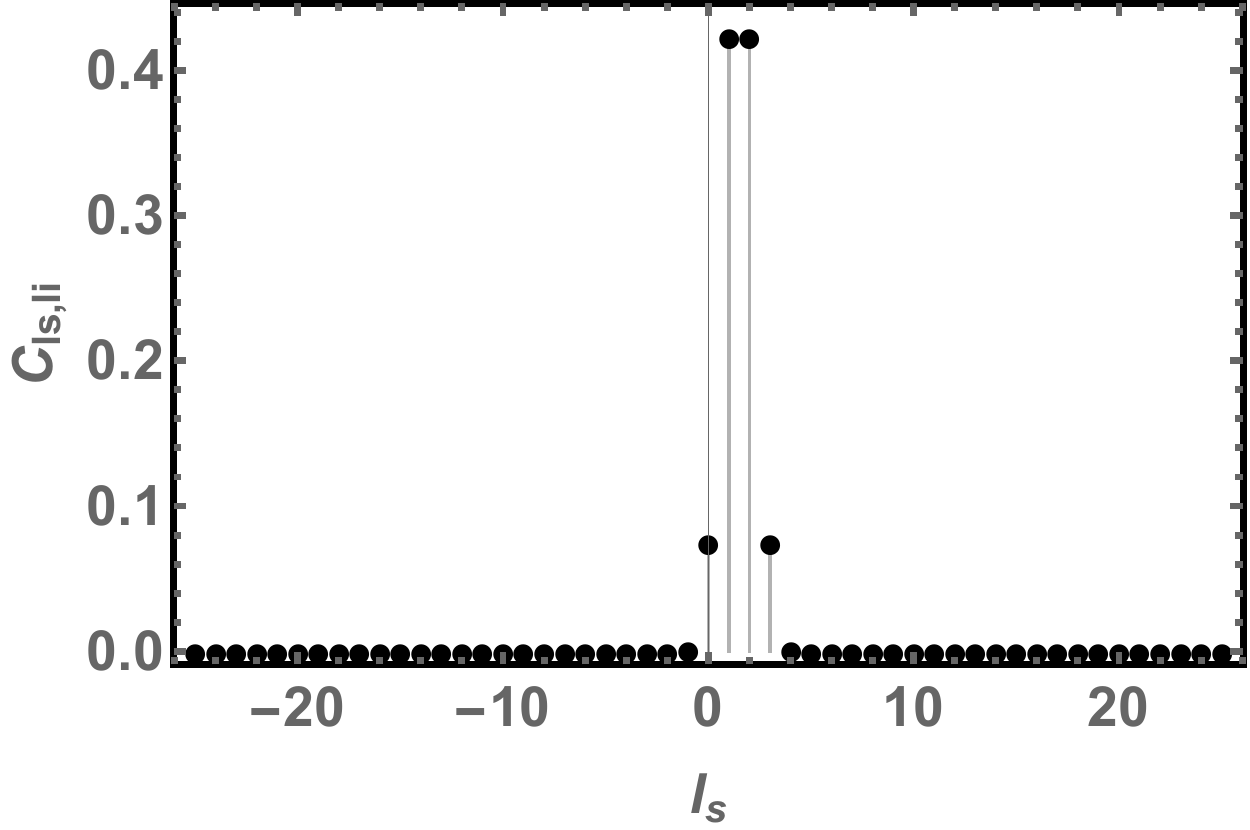}
\label{fig:label-h}}
\caption{OAM spectrum of the signal and idler photons for different pump OAM. Here the pump is in a POV mode and the signal and idler are projected to LG modes.}
\label{fig:PVSpectrum}
\end{figure}
In literature an ideal POV mode is written as
\begin{equation}
\label{perfectvortex}
\psi\left(r,\phi\right)=\frac{i^{l-1}}{k_r}\delta\left(r-r_0\right)\exp(il\phi)
\end{equation}
\noindent where $l$ is the azimuthal index as before, $r_0$ is the radius of the vortex in the transverse plane and $k_r$ is the radial wave vector. But realizing such a distribution is an experimental impossibility. In practice, a modified form of Eq. (\ref{perfectvortex}) can be generated by Fourier transforming an optical vortex in the Bessel-Gauss mode as shown in \cite{Vaity}. A Bessel-Gauss vortex mode has the distribution
\begin{equation}
\label{BGVortex}
\psi_l^{BG}\left(\rho,\theta\right) = J_l\left(k_r\rho\right)\exp\left(il\theta\right)\exp\left(\frac{\rho^2}{w^2}\right)
\end{equation}
\noindent where $J_l$ is the Bessel function of the first kind of order $l$ and $w$ is the beam waist of the gaussian envelope. This is Fourier transformed using a simple lens governed by the equation
\begin{eqnarray}
\label{fourier}
\psi\left(r,\phi\right) &=& \frac{k}{i2\pi f}\int_0^{\infty}\int_0^{2\pi}\psi\left(\rho,\theta\right)\nonumber \\
&\times & \exp\left(\frac{-ik}{f}\rho r\cos\left(\theta-\phi\right)\right)dV
\end{eqnarray}
\noindent where $f$ is the focal length of the lens used, $k=2\pi/\lambda$ is the magnitude of the total wave vector and $dV=\rho d\rho d\phi$. Eq. (\ref{BGVortex}) and Eq. (\ref{fourier}) can be used to derive an usable form of the POV mode as follows
\begin{equation}
\label{PVmode}
\psi_l^{POV}\left(r,\phi\right) = i^{l-1}\frac{w}{w_0}\exp\left(il\phi\right)\exp\left(-\frac{r^2+r_0^2}{w_0^2}\right)I_l\left(\frac{2r_0r}{w_0^2}\right)
\end{equation}
\noindent This is the same equation as Eq. (7) in \cite{Vaity}. Here, $r_0 = k_rf/k$ is the radius of the vortex ring of width $w_0$. An advantage of POV modes over other conventional vortex modes like LG and Bessel-Gaussian is that the diameter of the vortex ring is independent of the order of the vortex. The transverse distribution increases in size with increasing order of the vortex in case of LG and BG modes. But in case of POV modes, it remains invariant. This leads to better coupling efficiency with single mode fibers after phase flattening by using a computer generated hologram or a spatial light modulator. \\
Let us now consider the case where we use an optical vortex in a POV mode as the pump and project the signal and idler modes to LG modes given by Eq. (\ref{LGsimple}). We want to study how the projection of a POV mode to LG modes affects the OAM spectrum. In this case the overlap integral Eq. (\ref{coeff2}) takes the form
\begin{equation}
\label{PVcoeff}
C_{l_s,l_i}^{POV} = \int_V d^3r~ \psi^{POV}\left(r\right)\psi_{l_s}\left(r\right)\psi_{l_i}\left(r\right)
\end{equation}
\noindent where the indices are self-explanatory and have similar meaning as before. Using Eq. (\ref{PVmode}) and Eq. (\ref{LGsimple}) with correct indices in the above equation gives rise to the expansion coefficients $C_{l_s,l_i}^{POV}$. Proceeding as before we see from Fig. \ref{fig:PLGSpectrum} that this gives rise to a much narrower OAM spectrum. As expected, the most probable states are still the same as before with an added advantage that this decomposition does not give rise to the previously observed secondary maxima. The spectrum can be further narrowed by projecting the output modes to POV modes as can be seen from Fig. \ref{fig:PVSpectrum}. In this case the presence of higher order OAM modes is further reduced. This shows that projecting onto POV modes is the most efficient method to generate entangled photons in specific OAM modes with maximum probability as compared to the other two cases.
\newline It would be interesting to study how higher dimensional entanglement between the signal and idler photons are affected by the different OAM spectra which we do in the next section.

\section{Entanglement in the OAM basis}

In this section we study the entanglement between the signal and idler modes in the OAM basis. Rewriting Eq. (\ref{OAMBasis}) in a basis independent form, we obtain
\begin{equation}
\label{OAM}
\vert\psi\rangle=\sum_{l_s=-\infty}^{\infty}C_{l_s}^{l_p}\vert l_s\rangle\vert l_p-l_s\rangle
\end{equation}
\noindent where only OAM conservation $l_p=l_s+l_i$ is assumed. Since this is a pure state, the entanglement between the signal and idler modes can be faithfully quantified using the von Neumann entropy \cite{,Neumann,Nielsen}. For a bipartite state $\vert\psi_{AB}\rangle$ it is defined as
\begin{equation}
\label{entropydefinition}
S_B=-\sum_{k}\lambda_k \log_d\lambda_k
\end{equation}
\noindent where $\lambda_k$'s are the eigenvalues of the reduced density matrix $\rho_B=Tr_A~\left( \rho_{AB}\right)$. Here $Tr_A$ stands for partial trace operation that acts only on the $A$ part and $\rho_{AB}=\vert\psi_{AB}\rangle\langle\psi_{AB}\vert$ is the density matrix of the combined system. Here the logarithm is taken on base $d$ where $d$ stands for the dimensionality of the system. For example, $d=2$ for a two dimensional system. Therefore, an advantage of using the von Neumann entropy as a quantifier of entanglement is that it can be used even for higher dimensional entanglement. From Eq. (\ref{OAM}) it is easy to see that
\begin{equation}
\label{density}
\rho_{s,i}=\sum_{l_s,l_s'}C_{l_s}^{l_p}C_{l_s}^{l_p*}\vert l_s,l_p-ls\rangle\langle l_s',l_p-l_s'\vert
\end{equation}
\noindent from which it immediately follows that
\begin{equation}
\label{reduced}
\rho_i=\sum_{l_s} \vert C_{l_s}^{l_p}\vert^2~ \vert l_p-ls\rangle\langle l_p-l_s\vert
\end{equation}
\noindent where $\rho_i$ is the reduced density matrix for the idler mode. The above equation resembles the Schmidt decomposition for biphoton state in the OAM basis. It is the diagonal representation of the reduced density matrix of any quantum state. The expansion coefficients are the non-zero elements of this diagonal matrix. It has the interesting property that the Schmidt rank or the number of the coefficients in the decomposition can be directly used as an indicator of entanglement \cite{Sperling}. The reduced state is entangled iff the Schmidt rank is more than \emph{1}, else it is separable. The coefficients $\vert C_{l_s}^{l_p}\vert^2$ can then be used in place of $\lambda_k$ in Eq. (\ref{entropydefinition}) to calculate the entropy. These coefficients are the same as calculated from the overlap integral in the previous section.\\

\begin{table}[ht] 
\centering
\begin{tabular}{c c c c}
\hline\hline 
$l_p$ & LG$\to$LG,LG & POV$\to$LG,LG & POV$\to$POV,POV \\ [0.5ex] 
\hline 
0 & 1.8537 & 0.8850 & 0.3662 \\ 
1 & 2.4014 & 1.3921 & 0.8165 \\
2 & 2.7030 & 1.6016 & 1.0025 \\
3 & 2.9133 & 1.6998 & 1.1361 \\
4 & 3.0683 & 1.7469 & 1.2397 \\ [1ex] 
\hline 
\end{tabular}
\label{table:entropy} 
\caption{von Neumann entropy (in arbitrary units) of the idler mode calculated after tracing out the signal mode. $l_p$ is the OAM of the pump field.}
\end{table}

In Table I, we study the higher dimensional entanglement of the down converted photons for each of the decompositions considered earlier while studying the OAM spectrum. Please note that the results presented are in non-normalized arbitrary units. It is observed that maximum entanglement is seen to be present in case of LG $\to$ LG, LG for all values of the pump OAM as compared to the other two cases. The entanglement between the signal and idler photons is least when all the three modes, pump, signal and idler, are in POV modes. This is due to the decreasing probability of the occurence of higher order OAM states in the OAM basis expansion for each decomposition. Comparing Fig. \ref{fig:LGSpectrum}, Fig. \ref{fig:PLGSpectrum} and Fig. \ref{fig:PVSpectrum}, the probability of higher order OAM states is much higher when using a LG mode as the pump as well as projecting both signal and idler photons to LG modes. The probability decreases when the pump is replaced by a POV mode for the same projection scenario and is least when both the output photons are projected to POV modes.

\section{Conclusion}

In conclusion, we have studied the OAM spectrum of down converted photons for different pump modes. We considered LG and POV pump modes of various orders and looked at the resulting OAM spectrum of the SPDC photons for multiple projection scenarios. Specifically, we looked at those cases when both signal and idler photons are projected onto LG modes or POV modes. In order to compare the different projection scenarios on an equal footing, we considered only those LG modes for which the radial index in 0 since POV modes do not have any radial index. We found that projection of the output photons onto LG modes for a LG pump gave rise to a wide spectrum. It was noticed that higher order LG pump gives rise to secondary maxima. Replacing the pump with a POV mode removed this anomaly as well as narrowed down the spectrum. The narrowest spectrum was obtained when down converted photons were projected onto POV modes. In this case the the probability of occurence of higher order OAM states in the spectrum was reduced to a minimum and the spectrum was seen to be sharply peaked at those values of OAM for which the difference in OAM carried by signal and idler photon, $\vert l_s - l_i \vert$ is either 0 or 1. The narrowing of the spectrum, however, comes with a trade off. It leads to lower overall high-dimensional entanglement between signal and idler photons. So, to generate greater high-dimensional entangled state, a wider spectrum is required. This is found in the case of LG $\to$ LG, LG. But if a narrower spectrum is required (higher probability of the downconverted photons being in a specific OAM state), the best case scenario is the use of POV mode as pump as well as projection of signal and idler photons to POV modes.

\end{document}